\documentclass[reqno]{amsart} 
\usepackage{color}
\usepackage{amsaddr}
\usepackage{amsmath,amsthm}
\usepackage{amsfonts}
\usepackage[draft]{fixme}
\usepackage{tikz}
\usepackage{caption}
\usetikzlibrary{arrows}
\usepackage{comment}
\usepackage[all]{xy}

\usepackage{etoolbox}

\swapnumbers

\definecolor{myurlcolor}{rgb}{0.6,0,0}
\definecolor{mycitecolor}{rgb}{0,0,0.8}
\definecolor{myrefcolor}{rgb}{0,0,0.8}
\usepackage[pagebackref,bookmarks=false]{hyperref}
\hypersetup{colorlinks,
linkcolor=myrefcolor,
citecolor=mycitecolor,
urlcolor=myurlcolor}

\makeatletter
\def\@set@authors@addresses{\par
\skip@5\p@
\centering
\begingroup
\def\author##1{\par\vskip\skip@\MakeUppercase{##1}}%
\def\\{\protect\linebreak}%
\centering
\def\address##1##2{\par\addvspace\bigskipamount%
{\itshape\ignorespaces##2}%
 }%
\def\email##1##2{%
\@ifnotempty{##2}{, \ignorespaces{\ttfamily##2}}}%
\def\curraddr##1##2{}%
\def\urladdr##1##2{}%
\addresses
\endgroup
}
\makeatother

\newcommand{\define}[1]{{\bf \boldmath{#1}}}
\newcommand{\maps}{\colon}

\newcommand{\R}{\mathbb{R}}

\newcommand{\beq}{\begin{equation}}
\newcommand{\eeq}{\end{equation}}

\theoremstyle{plain}

\theoremstyle{definition}

\theoremstyle{remark}

\numberwithin{equation}{section}

\begin{document}



\title[Open Markov processes]{A Second Law for Open Markov Processes}

\author{Blake S. Pollard}
\address{Department of Physics and Astronomy\\
University of California \\
Riverside, CA 92521}
\email{bpoll002@ucr.edu}

\maketitle

\begin{abstract} In this paper we define the notion of an open Markov process. An open Markov process is a generalization of an ordinary Markov process in which populations are allowed to flow in and out of the system at certain boundary states. We show that the rate of change of relative entropy in an open Markov process is less than or equal to the flow of relative entropy through its boundary states. This can be viewed as a generalization of the Second Law for open Markov processes. In the case of a Markov process whose equilibrium obeys detailed balance, this inequality puts an upper bound on the rate of change of the free energy for any non-equilibrium distribution. 
\end{abstract}

\section{Introduction}

Markov processes are special cases of random walks or stochastic processes. Their utility stems from the fact that many otherwise intractable questions and concepts can be answered and explored using the framework of Markov processes. A Markov process can be viewed as a collection of states on which populations live. The `master equation' describes how populations hop from state to state. In this paper we define an open Markov process as one in which there are internal states, where the populations obey the master equation, and boundary states where populations do not obey the master equation because they interact with the external world. The state space of the composite system is the union of the boundary states and the internal states.

Often, the state space of a system interacting with its environment is given by the product of the state spaces of the system and the environment $S \times E$. Specifying a particular state corresponds to specifying the state of the system and the state of the environment. In the context of this article we consider a different viewpoint, where the state space of the composite system is given by the union of the internal and boundary states $S=I \cup B$. Thus a particle in the composite system can be in either an internal state or a boundary state. The interaction of the system with its environment is captured by the system's behavior at boundary states. 

One can visualize an open Markov process as a graph where the edges are labelled by positive real numbers. Each vertex is a `state' and the numbers attached to the edges are transition rates. Figure 1 shows an example of this graphical representation in which internal states are white and boundary states are shaded.

\begin{figure}[h]
\centering
\vbox{
\begin{tikzpicture}[->,>=stealth',shorten >=1pt,auto,node distance=3cm,
  thick,main node/.style={circle,fill=white!20,draw,font=\sffamily\Large\bfseries},terminal/.style={circle,fill=blue!20,draw,font=\sffamily\Large\bfseries}]]
  \node[main node](1) {a};
  \node[main node](2) [below right of=1] {b};
  \node[terminal](3) [below left of=1]  {c};
  \path[every node/.style={font=\sffamily\small}]
    (3) edge [bend left] node[above] {3} (1)
    (3) edge [bend right] node[above] {0.1} (2)
    (1) edge [bend left] node[right] {1.0} (2);  
\end{tikzpicture}
\caption{An open Markov process can be represented by a labelled graph. The numbers on each edge are transition rates. The white circles are internal states and the shaded circles are boundary states.}
\label{3state}
}
\end{figure}
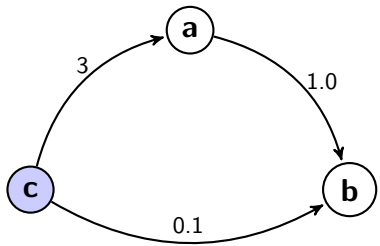

More precisely, an \define{open Markov process} is a triple $(V,H,B)$ where $V$ is a finite set of \define{states}, $H \maps \R^V \to \R^V$ is the \define{Hamiltonian} which is an \define{infinitesimal~stochastic} linear operator:  
\[ \begin{array}{ccc} \displaystyle{ \sum_{i \in V} H_{ij} = 0 }& \text{and}&  \displaystyle{ H_{ij} \geq 0 , i \neq j}.  \end{array} \]
The finite set $B \subseteq V$ is a subset of states called the \define{boundary states}. This also defines a set of \define{internal states}, $V-B$. Our dynamical variables are the \define{populations} of each state $p_i \in [0,\infty), i \in V$. The vector whose entries are the populations of each state at time $t$ we call the \define{population distribution}, $p(t) \in [0,\infty )^V$.  The off-diagonal entries $H_{ij}$ specify the rate at which population hops from state $j$ to state $i$. The Hamiltonian generates the time evolution of the populations at internal states via the \define{master equation}
\[ \frac{d p_i(t)}{dt} = \sum_{ j \in V} H_{ij} p_j(t) \ , \ i \in V-B .\]
In an open Markov process, populations at the boundary states do not obey the master equation, whereas populations at the internal states obey the master equation. An \define{ordinary Markov process} is an open Markov process, $(V,H,B)$, with all states being internal, i.e. $B= \{ \emptyset \} $. We simply write the pair $(V,H)$ for an ordinary Markov process. Note that in an ordinary Markov process the populations of all states satisfy the master equation. Kelly \cite{kelly2011reversibility} has described an analogy between open Markov processes and electrical circuits.


\subsection{Relative Entropy in Markov Processes}

Given two population distributions $p,q \in [0,\infty)^V$ the entropy of $p$ relative to $q$ or the \define{relative entropy} is given by:
\[ \displaystyle{ I(p,q) = \sum_{i \in V} p_i \ln \left( \frac{p_i}{q_i} \right). } \]
The relative entropy is sometimes referred to as the information gain or the Kullback--Leibler divergence \cite{kullback1951information}. Moran, Morimoto, and Csiszar proved that, in an ordinary Markov process, the entropy of any distribution relative to the equilibrium distribution is non-increasing \cite{csiszar1963informationstheoretische,moran1961entropy,morimoto1963markov}. Dupuis and Fischer proved that the relative entropy between any two distributions satisfying the master equation is non-increasing \cite{dupuis2012construction}. Merhav argues that the Second Law of thermodynamics can be viewed as a special case of the monotonicity in time of the relative entropy in Markov processes \cite{merhav2011data}.

The reason for using relative entropy instead of the usual Shannon entropy $ S(p) = -\sum_i p_i \ln(p_i) $ is that the usual entropy is not necessarily a monotonic function of time in Markov processes. If a Markov process has the uniform distribution as its equilibrium distribution, then the usual entropy will increase \cite{moran1961entropy}. A Markov process has the uniform distribution as its equilibrium distribution if and only if its Hamiltonian is \define{infinitesimal doubly stochastic}, meaning that both the columns and the rows sum to zero. Relative entropy is non-increasing even for Markov processes whose equilibrium distribution is not uniform \cite{cover1994processes}. This suggests the importance of a deeper underlying idea, that of the Markov ordering on the population distributions themselves; see \cite{gorban2010entropy} for details. For more information on reversibility and stochastic processes see \cite{alberti1982stochasticity,kelly2011reversibility}.

The goal of this paper is to study relative entropy in open Markov processes. We show that in an open Markov process $(V,H,B)$, if $p(t)$ and $q(t)$ obey the master equation at internal states then the rate of change of relative entropy satisfies the following inequality involving the behavior of the populations at the boundary states:
\[ \displaystyle { \frac{d}{dt} I( p(t), q(t) ) } \leq \displaystyle{ \sum_{i \in B} \frac{Dp_i}{Dt} \frac{\partial I}{\partial p_i} + \frac{Dq_i}{Dt} \frac{\partial I}{\partial q_i}  }.    \]
In this expression, $\frac{Dp_i}{Dt}$ is the \define{inflow} at the $i^{th}$ state, which is given by:
\[ \displaystyle{ \frac{Dp_i}{Dt}  = \frac{dp_i}{dt} - \sum_j H_{ij} p_j} . \] The inflow measures the amount by which the evolution of the population differs from that given by the master equation. The above inequality is our Second Law for open Markov processes. This inequality tells us that the rate of change of relative entropy in an open Markov process is less than or equal to the rate of change of relative entropy at the boundary. In Section 3 we derive this inequality.


\section{Composition of Open Markov Processes}

Part of the motivation for considering open Markov processes is to make precise the notion of composition of open Markov processes. One should be able to take two open Markov processes and combine them to get a new Markov process, where probability or population can now flow between the two original processes. This composition is accomplished by gluing two open Markov processes together along some set of boundary states. Since populations can flow in and out of an open Markov process through its boundary states, one needs to consider non-normalized measures.

Consider the two open Markov processes depicted in Figure 2. For concreteness let the states $ \{g,e,i\} $ correspond to a single atom in its ground, excited, and ionized states, respectively. For the purposes of this example, let us assume we are in a regime where environmentally-induced decoherence allows us to treat the process of an atom transitioning between states as a Markov process. 

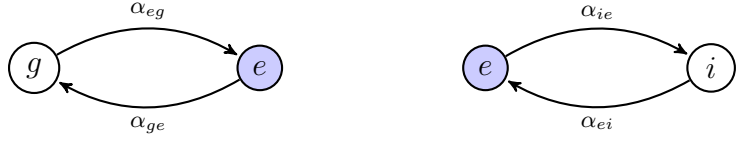
\begin{figure}[h] 
\begin{center}
\begin{tikzpicture}[->,>=stealth',shorten >=1pt,auto,node distance=3cm,
  thick,main node/.style={circle,fill=white!20,draw,font=\sffamily\Large\bfseries},terminal/.style={circle,fill=blue!20,draw,font=\sffamily\Large\bfseries}]]
  \node[main node](1) {$g$};
  \node[terminal](2) [right of=1] {$e$};
  \node[terminal](3) [right of=2]  {$e$};
  \node[main node](4) [right of=3] {$i$};
  \path[every node/.style={font=\sffamily\small}]
    (3) edge [bend left] node[above] {$\alpha_{ie}$} (4) 
    (4) edge [bend left] node[below] {$\alpha_{ei}$} (3)
    (1) edge [bend left] node[above] {$\alpha_{eg}$} (2)
    (2) edge [bend left] node[below] {$\alpha_{ge}$} (1);  
\end{tikzpicture}
\end{center}
\caption{Two open Markov processes modeling the transitions of an atom between ground and excited states and between excited and ionized states.}
\label{2processes}
\end{figure}

In order to capture the possibility that an atom transitions from its ground state to an excited state and then becomes ionized we compose the two open Markov processes to give a new Markov process, depicted in Figure 3. In this example we suppose that in the process of ionization, an atom always passes through an excited state and vice-versa.

\begin{figure}[h!]
\begin{center}
\begin{tikzpicture}[->,>=stealth',shorten >=1pt,auto,node distance=3cm,
  thick,main node/.style={circle,fill=white!20,draw,font=\sffamily\Large\bfseries},terminal/.style={circle,fill=blue!20,draw,font=\sffamily\Large\bfseries}]]
  \node[main node](1) {$g$};
  \node[main node](2) [right of=1] {$e$};
  \node[main node](3) [right of=2]  {$i$};
  \path[every node/.style={font=\sffamily\small}]
    (2) edge [bend left] node[above] {$\alpha_{ie}$} (3) 
    (3) edge [bend left] node[below] {$\alpha_{ei}$} (2)
    (1) edge [bend left] node[above] {$\alpha_{eg}$} (2)
    (2) edge [bend left] node[below] {$\alpha_{ge}$} (1);  
\end{tikzpicture}
\end{center}
\caption{Composition of the two open Markov processes in Figure 2 results in an ordinary Markov process. }
\label{composite}
\end{figure}
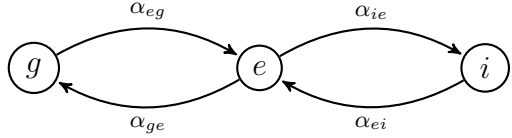

Notice that after composition, the state labelled $e$ becomes an internal state. For this particular example the result of composition is an open Markov process with no boundary states, which is an ordinary Markov process. In general, composition of open Markov processes may result in another open Markov process, i.e. after composition there may be a non-empty set of boundary states.

The master equation for this Markov process is
\[ \begin{array}{ccc} 
\displaystyle{ \frac{dp_g}{dt} } &=& -\alpha_{eg} p_g + \alpha_{ge} p_e \\ \\
\displaystyle{ \frac{dp_e}{dt} }&=& \alpha_{eg} p_g - \alpha _{ge} p_e - \alpha_{ie} p_e + \alpha_{ei} p_i \\ \\
\displaystyle{ \frac{dp_i}{dt} } &=& -\alpha_{ei} p_i + \alpha_{ie} p_e .

\end{array}
\]
In this case we have $\frac{d}{dt} \left( p_g + p_e + p_i \right) = 0 $ and the total population is conserved in time. If we consider only the states of one of the open Markov processes in Figure 2, then population can flow in and out through the boundary and we have $ \frac{d}{dt} \left( p_g + p_e \right) \neq 0$; similarly for $p_e$ and $p_i$. Because we would like to analyze the behavior of relative entropy within the subsystems themselves, we need to work with non-normalized measures.


\subsection{Relative Entropy is Subadditive Under Composition}

Suppose we have an ordinary Markov process $(V,H)$, which is the composite of two open Markov processes $(V_1, H_1, B_1)$ and $(V_2, H_2, B_2)$. We consider the case, as in Figures 2 and 3, when $V = V_1 \cup V_2$ and $ B_2=B_1=V_1 \cap V_2$. Given two population distributions $p$ and $q$ on $V$, let us define the following notation for relative entropy:
\[ \displaystyle{I_V( p, q ) = \sum_{i \in V} p_i \ln \left( \frac{p_i}{q_i} \right) }. \]
 Using this notation, we write the relative entropy of the composite as,
\[ \displaystyle{ I_{V_1 \cup V_2}(p,q) = I_{V_1}(p,q) + I_{V_2}(p,q) - I_{V_1 \cap V_2}(p,q) }. \]
The third term comes from the fact that the contributions to the relative entropy from the boundary states are counted in both the sum over $V_1$ and the sum over $V_2$.


\section{The Second Law for Open Markov Processes}

In this section we show that the rate of change of relative entropy in an open Markov process is less than or equal to the relative entropy flowing through its boundary states. We use the fact that relative entropy is non-increasing in an ordinary Markov process. 


\subsection{Relative Entropy Change in Ordinary Markov Processes}

Given a Markov process $(V,H)$ and two population distributions $p(t),q(t) \in \R^V$, each of which are solutions to the master equation, the entropy of $p(t)$ relative to $q(t)$ is
\[ \displaystyle{ I (p(t),q(t) ) =  \sum_i p_i(t) \ln \left( \frac{p_i(t)}{q_i(t)} \right). } \]
Following Dupuis and Fischer \cite{dupuis2012construction}, we can see that relative entropy is non-increasing for Markov processes:
\[ \begin{array}{ccl} \displaystyle{ \frac{d I( p(t), q(t) ) } {dt} } &=& \displaystyle{ \frac{d}{dt} \sum_i p_i \ln \left( \frac{p_i}{q_i} \right)  } \\ \\
&=& \displaystyle{  \sum_i \frac{dp_i}{dt} \ln \left( \frac{p_i}{q_i} \right) +\sum_i q_i \frac{d}{dt} \left( \frac{p_i}{q_i} \right) } \\ \\
&=& \displaystyle{  \sum_i \left[ \sum_j H_{ij} p_j \ln \left( \frac{p_i}{q_i} \right) +\sum_j H_{ij}p_j - \sum_j \frac{p_i}{q_i} H_{ij} q_j \right] } \\ \\
&=& \displaystyle{  \sum_i \left[ \sum_{ j \neq i} H_{ij} p_j \left[ \ln \left( \frac{p_i}{q_i} \right) -\frac{p_i q_j}{p_j q_i} + 1 \right] + H_{ii}p_i \ln \left( \frac{p_i}{q_i} \right) \right] } \\ \\ 
&=& \displaystyle{  \sum_i \sum_{ j \neq i} H_{ij} p_j \left[ \ln \left( \frac{p_i}{q_i} \right) -\frac{p_i q_j}{p_j q_i} + 1 \right] + \sum_j H_{jj} p_j \ln \left( \frac{p_j}{q_j} \right) } \\ 
\end{array} \]
\[ \begin{array}{ccl}
&=& \displaystyle{  \sum_i \sum_{j \neq i} H_{ij} p_j \left[ \ln \left( \frac{p_i}{q_i} \right) - \frac{p_i q_j}{p_j q_i} + 1 \right] - \sum_j \sum_{i \neq j} H_{ij} p_j \ln \left( \frac{p_j}{q_j} \right) } \\ \\
&=& \displaystyle{  \sum_i \sum_{j \neq i} H_{ij} p_j \left[ \ln \left( \frac{p_i q_j}{q_i p_j} \right) - \frac{p_i q_j}{q_i p_j} +1 \right] } \\ \\
&\leq& 0. \end{array} \]
The last line follows from the fact that $H_{ij} \geq 0$ for $i \neq j$ along with the fact that the term in the brackets $\ln(x)-x+1$ 
is everywhere negative except at $x=1$ where it is zero. As $q_i \rightarrow 0$ for some $i \in V$, the rate of change of relative entropy tends towards negative infinity. One has to allow infinity as a possible value for relative entropy and negative infinity as a possible value for its first time derivative, in which case the above inequality still holds.  Thus, we conclude that for any ordinary Markov process,
\[ \displaystyle{ \frac{d}{dt} I(p(t),q(t)) \leq 0}.  \]
This inequality is the continuous-time analog of the generalized data processing lemma \cite{cohen1993relative,cohenmajorization}. It holds for any two, non-normalized, population distributions $p$ and $q$. 

Since the Second Law of Thermodynamics says that entropy never decreases, it may seem odd that relative entropy never increases. However, if the reference distribution $q$ is taken to be the uniform distribution $q_i  = c$ for all $i$ and for some constant $c$, then the relative entropy becomes
\[ \displaystyle{ I(p,q) = \sum_i p_i \ln(p_i) - \sum_i p_i \ln(c). } \]
If $\sum_i p_i $ is constant, then for $q$ uniform, the relative entropy equals the negative of the usual entropy minus a constant. Thus the above calculation for $ \frac{d I(p,q) }{dt}$ gives the usual Second Law.  


\subsection{Relative Entropy Change in Open Markov Processes}

Now we calculate the rate of change of relative entropy in an open Markov process $(V,H,B)$.
Recall that the inflow at the $i^{th}$ vertex is given by
\[ \displaystyle{ \frac{D p_i}{Dt} = \frac{dp_i}{dt}-\sum_{j \in V} H_{ij} p_j . } \]
Note that the inflow is zero for internal states as the master equation holds at internal states. Also note the following relations:
\[ \displaystyle{ \frac{\partial I(p,q) }{\partial p_i} = \sum_i \left( \ln \left(\frac{p_i}{q_i} \right) +1 \right) } \]
and
\[ \displaystyle{ \frac{\partial I(p,q) } { \partial q_i} = -\sum_i \frac{p_i}{q_i}. } \]
Taking the time derivative of the relative entropy we obtain
\[  \begin{array}{ccl} \displaystyle{ \frac{d}{dt} I( p(t), q(t) )  } &=& \displaystyle{ \sum_{ i \in V} \frac{dp_i}{dt} \left[ \ln \left( \frac{p_i}{q_i} \right) +1 \right] - \sum_{i \in V} \frac{p_i}{q_i} \frac{dq_i}{dt} } \\ \\
&=& \displaystyle{ \sum_{i \in V-B} \sum_{j \in V} H_{ij} p_j \left[ \ln \left( \frac{p_i}{q_i} \right) + 1  - \frac{p_i q_j}{q_i p_j} \right] } \\ \\
& & \displaystyle{ + \sum_{i \in B} \left[ \frac{dp_i}{dt} \left[ \ln \left( \frac{p_i}{q_i} \right)+1\right] - \frac{p_i}{q_i} \frac{dq_i}{dt} \right] }. \\ \\ \end{array} \]
In the last step we separated the contributions from internal and boundary states and used the master equation for the internal states. Now let us add and subtract terms so that the first term corresponds to the rate of change of relative entropy for a Markov process with no boundary states:
\[ \begin{array}{ccl} \displaystyle{ \frac{d}{dt} I( p(t), q(t) ) } 
&=& \displaystyle{ \sum_{i \in V} \sum_{j \in V} H_{ij} p_j \left[ \ln \left( \frac{p_i}{q_i} \right) + 1  - \frac{p_i q_j}{q_i p_j} \right] } \\ \\
& & \displaystyle{ + \sum_{i \in B} \sum_{j \in V} \left( \frac{dp_i}{dt} - H_{ij}p_j \right) \left( \ln \left( \frac{p_i}{q_i} \right) + 1\right) } \\ \\
& & \displaystyle{ - \sum_{i \in B} \sum_{j \in V}  \left(\frac{dq_i}{dt} - H_{ij} q_j \right) \frac{p_i}{q_i} }.
\end{array} \]
The first term is the rate of change of relative entropy for an ordinary Markov process, which is less than or equal to zero. Therefore, we have
\[ \begin{array}{ccl} \displaystyle{ \frac{d}{dt} I ( p(t),q(t) ) } 
&\leq& \displaystyle{  \sum_{i \in B} \sum_{j \in V}  \left( \frac{dp_i}{dt} - H_{ij}p_j \right) \left( \ln \left( \frac{p_i}{q_i} \right) + 1\right)  } \\
& & \displaystyle{ - \sum_{i \in B} \sum_{j \in V} \left(\frac{dq_i}{dt} - H_{ij} q_j \right) \frac{p_i}{q_i}  }. \end{array} \]
We can write this more compactly as
\[ \begin{array}{ccl} \displaystyle{ \frac{d}{dt} I ( p(t),q(t) ) }
&\leq& \displaystyle{  \sum_{ i \in B} \frac{Dp_i}{Dt}\frac{\partial I}{\partial p_i} + \frac{Dq_i}{Dt}\frac{\partial I}{\partial q_i} } \end{array}. \]
This gives a version of the Second Law that holds for open Markov processes. One can see that this result reduces to the usual Second Law for an ordinary Markov process, where all states are internal and there are no boundary states. 


\section{Thermodynamic Interpretation}
The possibility of increasing relative entropy is a generic feature of interacting systems. For a closed system, relative entropy can increase within a particular subsystem, but as was shown in section 3.1 this increase will always be compensated by a decrease elsewhere in the system. This is analogous to the case of entropy in thermodynamics. The generalization of the Second Law to the type of open systems described in this article can be applied to non-equilibrium thermodynamic systems where external forcings at boundary states maintain the system out of equilibrium. 

Consider the case of an ordinary Markov process whose equilibrium distribution $q$ satisfies detailed balance, $H_{ij}q_j = H_{ji}q_i$. If to each state we associate an energy $E_i$, then we can write the $q_i$'s as Gibbs states
\[ q_i = \frac{ e^{-\beta E_i} }{\mathcal{Z}},\]  where $\beta = \frac{1}{T}$ is the inverse temperature in units where Boltzmann's constant is equal to one. The partition function $\mathcal{Z}=e^{-\beta F[q]}$ can be used to adjust the normalization of $q$. Here, $F[q] = \langle E \rangle_q -TS(q)$ is the equilibrium free energy where $S(q) = -\sum_i q_i \ln{q_i}$ is the Shannon entropy. The entropy of a non-equilibrium state $p(t)$ relative to the equilibrium $q$ gives
\[ I(p(t), q) = \sum_i p_i(t) \ln \left( \frac{p_i(t)}{q_i} \right), \]
which can be written as 
\[ I(p(t),q) = -S(p) + \beta \langle E \rangle_{p(t)} - \beta  F[q]. \] 
If we define the free energy of the non-equilibrium distribution $p$ as $F[p] = \langle E \rangle_p - TS(p)$ we have that 
 \[ I(p(t),q) = \frac{ F[p(t)] - F[q] }{T}.\]
Thus in the case where $q$ is an equilibrium distribution of the ordinary Markov process satisfying detailed balance then the relative entropy $I(p(t),q)$ is simply the amount by which the free energy of $p(t)$ exceeds the equilibrium free energy, divided by the temperature. 

Since $q$ is an equilibrium of the ordinary Markov process we have that $\frac{dq_i}{dt} = \sum_j H_{ij}q_j = 0 \ \forall \ i \in V$. In this case our inequality for open Markov processes reads
\[ \frac{d}{dt} F[p(t)] \leq T\sum_{ i \in B} \frac{Dp_i}{Dt}\frac{\partial I}{\partial p_i}. \] 
If the $p_i$ were to obey the master equation at all states the right-hand side of this expression would vanish, indicating that the free energy of $p_i(t)$ approaches the equilibrium free energy. For a system held out of equilibrium by some external forcings along its boundary, this inequality says that the rate of change of free energy cannot exceed the temperature times the rate of inflow of relative entropy.


\section{Conclusion}
The desire to view a complicated Markov processes as being built up from the composition of a number of subprocesses led us to introduce the concept of an open Markov process, where the populations at certain boundary states do not satisfy the master equation. We described a method for composing two open Markov processes to get a new Markov process and showed that relative entropy is subadditive under this composition. Since populations are allowed to flow in and out of an open Markov process, the total population is not conserved, necessitating the use of non-normalized measures. We then analyzed the behavior of relative entropy in these open processes. Working with non-normalized populations allows the relative entropy to take on negative values. In this paper we have shown that relative entropy is non-increasing even for non-normalized population distributions.

Using this result, we were able to derive an inequality bounding the rate of relative entropy production for open processes. We provided a thermodynamic interpretation of this inequality in the special case where the equilibrium distribution of the Markov process satisfies detailed balance. Open Markov processes provide a framework for describing population distributions which deviate from the master equation at certain states. Many natural systems are only approximately Markovian. The Second Law for open Markov processes quantifies such deviations by giving an explicit formula bounding the rate of relative entropy generation. 


\subsection*{Acknowledgements}
I am indebted to John C. Baez for his guidance as well as for numerous conversations which helped shape this paper. I thank the Centre for Quantum Technologies and everyone there for their hospitality during my visit. I also thank the FQXi for funding my visit.

\end{document}